\documentclass[a4paper]{article}
\usepackage{graphicx}
\usepackage{twocolceurws}
\usepackage{color}
\usepackage{float}
\usepackage{booktabs}
\usepackage{svg}
\usepackage{url}

\title{I Stand With You: Using Emojis to Study Solidarity in Crisis Events}
\author{Sashank Santhanam\textsuperscript{1}\textsuperscript{*}, Vidhushini Srinivasan\textsuperscript{1}, Shaina Glass\textsuperscript{2}, Samira Shaikh\textsuperscript{1}}
\institution{ \textsuperscript{1}{Department of Computer Science, \textsuperscript{2}Department of Psychology}\\University of North Carolina at Charlotte\\ \textsuperscript{*}{ssantha1@uncc.edu}
}
\begin{document}
\maketitle
\begin{abstract}
We study how emojis are used to express solidarity in social media in the context of two major crisis events - a natural disaster, Hurricane  Irma in 2017 and terrorist attacks that occurred in November 2015 in Paris. Using annotated corpora, we first train a recurrent neural network model to classify expressions of solidarity in text. Next, we use these expressions of solidarity to characterize human behavior in online social networks, through the temporal and geospatial diffusion of emojis. Our analysis reveals that emojis are a powerful indicator of sociolinguistic behaviors  (solidarity) that are exhibited on social media as the crisis events unfold.    
\end{abstract}

\section{Introduction}

The collective enactment of online behaviors, including prosocial behaviors such as solidarity, has been known to directly affect political mobilization and social movements \cite{tufekci2014social,fenton2008mediating}. Social media, due to its increasingly pervasive nature, permits a sense of immediacy \cite{giddens2013consequences} - a notion that produces high degree of identification among politicized citizens of the web, especially in response to crisis events \cite{fenton2008mediating}. Furthermore, the multiplicity of views and ideologies that proliferate on Online Social Networks (OSNs) has created a society that is increasingly fragmented and polarized \cite{del2016echo,sunstein2018republic}. Prosocial behaviors like solidarity then become necessary and essential in overcoming ideological differences and finding common ground \cite{bauman2013postmodernity}, especially in the aftermath of crisis events (e.g. natural disasters). Recent social movements with a strong sense of online solidarity have had tangible offline (real-world) consequences, exemplified by movements related to \textit{\#BlackLivesMatter}, \textit{\#MeToo} and \textit{\#NeverAgain} \cite{de2016social,bureau_2018}. There is thus a pressing need to understand how solidarity is expressed online and more importantly, how it drives the convergence of a global public in OSNs. 

Furthermore, research has shown that emoticons and emojis are more likely to be used in socio-emotional contexts \cite{derks2007emoticons} and that they may serve to clarify the message structure or reinforce the message content \cite{markman2017pragmatic,donato2017investigating}. Riordan \cite{riordan2017communicative} found that emojis, especially non-face emojis, can alter the reader's perceived affect of messages. While research has investigated the use of emojis over communities and cultures \cite{barbieri2016cosmopolitan,ljubevsic2016global} as well as how emoji use mediates close personal relationships \cite{kelly2015characterising}, the systematic study of emojis as indicators of human behaviors in the context of social movements has not been undertaken. We thus seek to understand how emojis are used when people express behaviors online on a global scale and what insights can be gleaned through the use of emojis during crisis events. Our work makes two salient contributions:
\begin{itemize}
\itemsep0em
    \item We make available two large-scale corpora\footnote{\url{https://github.com/sashank06/ICWSM_Emoji}}, annotated for expressions of solidarity using mutiple annotators and containing a large number of emojis, surrounding two distinct crisis events that vary in time-scales and type of crisis event.
    \item A framework and software for analyzing of how emojis are used to express prosocial behaviors such as solidarity in the online context, through the study of temporal and geospatial diffusion of emojis in online social networks.
\end{itemize}
We anticipate that our approach and findings would help advance research in the study of online human behaviors and in the dynamics of online mobilization. 

\section{Related Work}
\textbf{Defining Solidarity:} We start by defining what we mean by solidarity. The concept of solidarity has been defined by scholars in relation to complementary terms such as ``community spirit or mutual attachment, social cooperation or charity'' \cite{bayertz1999four}. In our work, we use the definition of \textit{expressional solidarity} \cite{taylor2015solidarity}, characterized as individuals expressing empathy and support for a group they are not directly involved in (for example, expressing solidarity for victims of natural disasters or terrorist attacks). 

\textbf{Using Emojis to Understand Human Behavior:} With respect to research on expressional solidarity, Herrera et al. found that individuals were more outspoken on social media after a tragic event \cite{herrera2015solidarity}. They studied solidarity in tweets spanning geographical areas and several languages relating to a terrorist attack, and found that hashtags evolved over time correlating with a need of individuals to speak out about the event. However, they did not investigate the use of emojis in their analysis. 

Extant research on emojis usage has designated three categories, that are 1) function: when emojis replace a conjunction or prepositional word; 2) content: when emojis replace a noun, verb, or adjective; and 3) multimodal: when emojis are used to express an attitude, the topic of the message or communicate a gesture \cite{na2017varying}. \cite{na2017varying} found that the multimodal category is the most frequently used; and we contend that emojis used in the multimodal function may also be most likely to demonstrate solidarity. Emojis are also widely used to convey sentiment \cite{hu2017spice}, including strengthening expression, adjusting tone, expressing humor, irony, or intimacy, and to describe content, which makes emojis (and emoticons) viable resources for sentiment analysis \cite{jibril2013relevance,novak2015sentiment,pavalanathan2015emoticons}.  We use sentiment of emojis to study the diffusion of emojis across time and region.

To the best of our knowledge, no research to date has described automated models of detecting and classifying solidarity expressions in social media. In addition, research on using such models to further investigate how human behavior, especially a prosocial behavior like solidarity, is communicated through the use of emojis in social media is still nascent. Our work seeks to fill this important research gap.

\section{Data Collection}
Our analysis is based on social media text surrounding two different crisis attacks: Hurricane Irma in 2017 and terrorist attacks in Paris, November 2015. We begin this section by briefly describing the two corpora. 

\textbf{Irma Corpus:} Hurricane Irma was a catastrophic Category 5 hurricane and was one of the strongest hurricanes ever to be formed in the Atlantic\footnote{https://tinyurl.com/y843u5kh}. The storm caused massive destruction over the Caribbean islands and Cuba before turning north towards the United States. People took to social media to express their thoughts along with tracking the progress of the storm. To create our Irma corpus, we used Twitter streaming API to collect tweets with mentions of the keyword ``irma'' starting from the time Irma became an intense storm (September 6\textsuperscript{th}, 2017) and until the storm weakened over Mississippi on September 12\textsuperscript{th}, 2017 resulting in a corpus of $>$16MM tweets. 

\textbf{Paris Corpus:} Attackers carried out suicide bombings and multiple shootings near cafes and the Bataclan theatre in Paris on November 13\textsuperscript{th}, 2015. More than 400 people were injured and over a hundred people died in this event\footnote{https://tinyurl.com/pb2bohv}. As a reaction to this incident, people all over the world took to social media to express their reactions. To create our Paris corpus, we collected $>$2MM tweets from 13\textsuperscript{th} November, 2015 to 17\textsuperscript{th} November, 2015 containing the word ``paris'' using the Twitter GNIP service\footnote{https://tinyurl.com/y8amahe6}.

\textbf{Annotation Procedure}
We performed distance labeling \cite{mintz2009distant} by having two trained annotators assign the most frequent hashtags in our corpus with one of three labels (``Solidarity'' (e.g. \textit{\#solidaritywithparis, \#westandwithparis, \#prayersforpuertorico}), ``Not Solidarity'' (e.g. \textit{\#breakingnews, \#facebook}) and ``Unrelated/Cannot Determine'' (e.g. \textit{\#rebootliberty, \#syrianrefugees}). Using the hashtags that both annotators agreed upon ($\kappa$ $>$ 0.65, which is regarded as an acceptable agreement level) \cite{sim2005kappa}, we filtered tweets that were annotated with conflicting hashtags from both corpora, as well as retweets and duplicate tweets. Table \ref{descriptives} provides the descriptive statistics of the original (not retweets), non-duplicate tweets, that were annotated as expressing solidarity and not solidarity based on their hashtags that we used for further analysis.

\begin{table}[H]
\centering
\caption{Descriptive statistics for crisis event corpora}
\label{descriptives}
\begin{tabular}{@{}llll@{}}
\toprule
       \# of Tweets  & \textbf{Solidarity} & \textbf{Not Solidarity} & \textbf{Total} \\ \midrule
\textbf{Irma}  & 12000               & 81697            & 93697          \\
\textbf{Paris} & 20465               & 29874            & 50339 \\
\hline
\end{tabular}
\end{table}

\section{The Emojis of Solidarity}

The main goal of this article is to investigate how individuals use emojis to express a prosocial behavior, in this case, solidarity, during crisis events. Accordingly, we outline our analyses in the form of research questions (RQs) and the resulting observations in the sections below.

\textbf{RQ1: How useful are emojis as features in classifying expressions of solidarity?}

After performing manual annotation of the two corpora, we trained two classifiers for detecting solidarity in text. We applied standard NLP pre-processing techniques of tokenization, removing stopwords and lowercasing the tweets. We also removed hashtags that were annotated from the tweets. Class balancing was used in all models to address the issue of majority class imbalance (count of Solidarity vs. Not Solidarity tweets). 

\textbf{Baseline Models:} We used Support Vector Machine (SVM) with a linear kernel and 10 fold cross validation to classify tweets containing solidarity expressions. For the baseline models, we experimented with three variants of features including (a) word bigrams, (b) TF-IDF \cite{manning2008info}, (c) TF-IDF+Bigrams. 

\textbf{RNN+LSTM Model:} We built a Recurrent Neural Network(RNN) model with Long Short-Term Memory (LSTM) \cite{hochreiter1997long} to classify social media posts into Solidarity and Not Solidarity categories. The embedding layer of the RNN is initialized with pre-trained GloVe embeddings \cite{pennington2014glove} and the network consists of a single LSTM layer. All inputs to the network are padded to uniform length of 100. Table \ref{hyperparameters} shows the hyperparameters of the RNN model.

Table \ref{results} shows the accuracy of the baseline and RNN+LSTM models in classifying expressions of solidarity from text, where the RNN+LSTM model with emojis significantly outperforms the Linear SVM models in both Irma and Paris corpora.

\begin{table}[ht]
\centering
\caption{RNN+LSTM model hyperparameters}
\label{hyperparameters}
\begin{tabular}{|l|l|}
\hline
\textbf{Hyperparameters}     & \textbf{Value} \\ \hline
Batch Size          & 25 \\ \hline
Learning Rate & 0.001               \\ \hline
Epochs   & 20             \\ \hline
Dropout       & 0.5 \\ \hline
\end{tabular}
\end{table}
\begin{table}[ht]
\centering
\caption{Accuracy of the baseline SVM models and RNN+LSTM model} 
\label{results}
\begin{tabular}{|l|l|l|}
\hline
\textbf{Accuracy} & \textbf{Irma}    & \textbf{Paris}  \\ \hline
RNN+LSTM (with emojis)       & \textbf{93.5\%} & \textbf{86.7\%}  \\ \hline
RNN+LSTM (without emojis)        & 89.8\% & 86.1\%  \\ \hline
TF-IDF            & 85.71\%         &    75.72\%        \\ \hline
TF-IDF + Bigrams  & 82.62\%         & 76.98\%           \\ \hline
Bigrams only           & 79.86\%        & 75.24\%    \\ \hline

\end{tabular}
\end{table}

\textbf{RQ2: Which emojis are used in expressions of solidarity during crisis events and how do they compare to emojis used in other tweets?}

To start delving into the data, in Table \ref{emoji_descriptives} we show the total number of emojis in each dataset. We observe that the total number of emojis in the tweets that express solidarity (using ground-truth human annotation) is greater than the emojis in not solidarity tweets, even though the number of not solidarity tweets is greater than the solidarity tweets in both crisis events (c.f. Table \ref{descriptives}). We also observe that count of emojis is greater in the Irma corpus than in the Paris corpus, even though the number of solidarity tweets is smaller in the Irma corpus. One reason for this could be that the Hurricane Irma event happened in 2017 when predictive emoji was a feature on platforms, while the Paris event occurred in 2015 when such functionality was not operational.  

\begin{table}[H]
\centering
\caption{Total number of emojis in each dataset}
\label{emoji_descriptives}
\begin{tabular}{@{}llll@{}}
\toprule
       \# of Emojis  & \textbf{Solidarity} & \textbf{Not Solidarity} & \textbf{Total}\\ \midrule
\textbf{Irma}  & 26197               & 25904            & 52101          \\
\textbf{Paris} & 24801               & 12373            & 37174 \\
\hline
\end{tabular}
\end{table}

\begin{table}[H]
\centering
\caption{Top ten emojis by frequency and their counts in Irma and Paris corpora}
\label{emoji_ranks}
 \begin{tabular}{|l|l|l|l|l|} 
 \hline
 \textbf{Rank} & \vtop{\hbox{\strut \textbf{Irma}}\hbox{\strut \textbf{Sol.}}} & \vtop{\hbox{\strut \textbf{Irma}}\hbox{\strut \textbf{Not Sol.}}} & \vtop{\hbox{\strut \textbf{Paris}}\hbox{\strut \textbf {Sol.}}} & \vtop{\hbox{\strut \textbf{Paris}}\hbox{\strut \textbf {Not Sol.}}} \\  
 \hline\hline
 1 &  \includegraphics[height=1em]{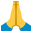} 6105 & \includegraphics[height=1em]{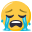} 2098 & 
 \includegraphics[height=1em]{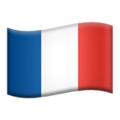} 5376 & \includegraphics[height=1em]{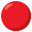} 2878 \\ 
 \hline
 2  & \includegraphics[height=1em]{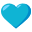} 2336 & \includegraphics[height=1em]{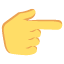} 1827 & \includegraphics[height=1em]{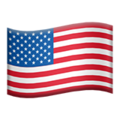} 2826 & \includegraphics[height=1em]{images/1f1eb-1f1f7_rect.png} 1033\\
 \hline
 3  & \includegraphics[height=1em]{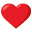} 1977 & \includegraphics[height=1em]{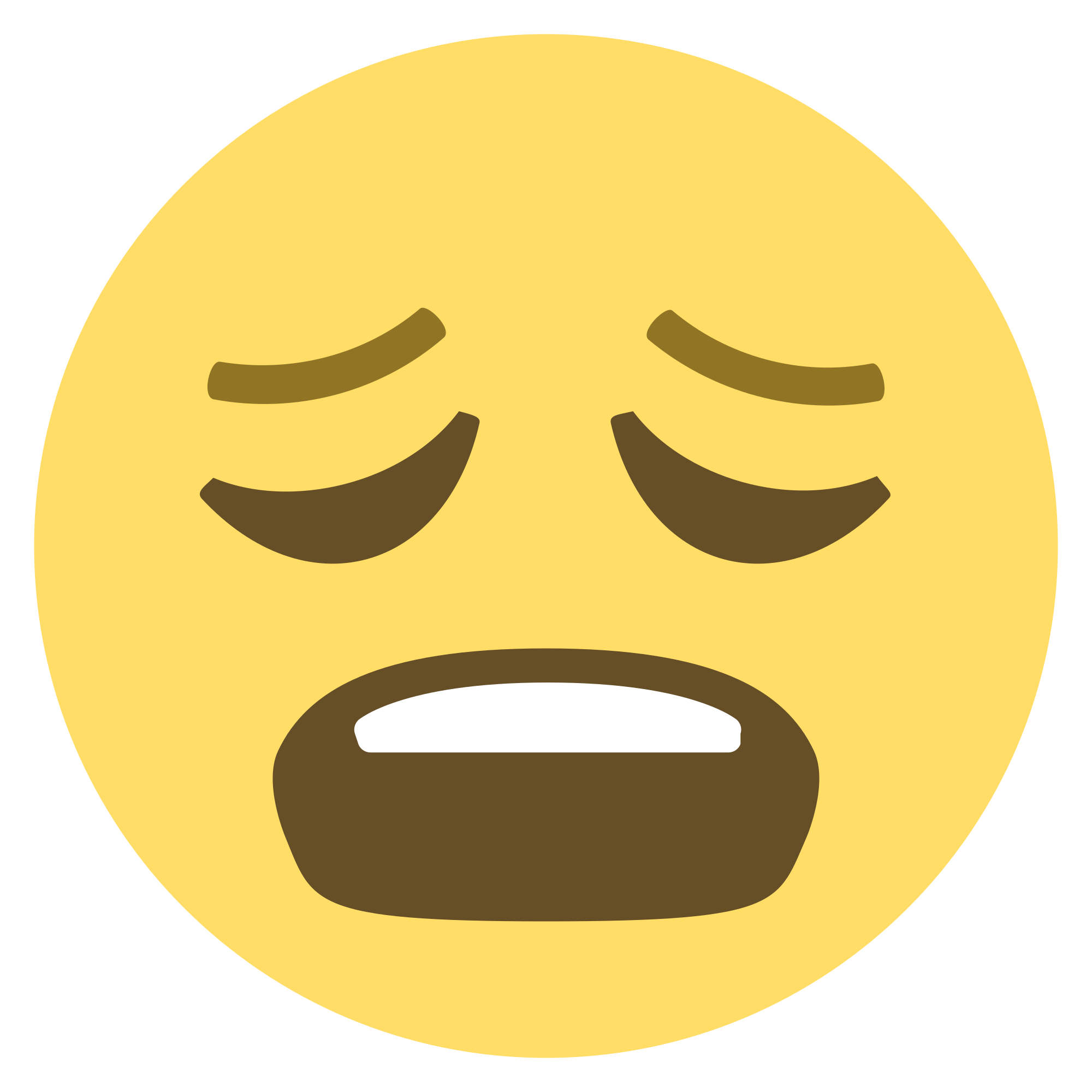} 1474 & \includegraphics[height=1em]{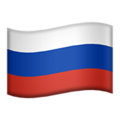} 2649 & \includegraphics[height=1em]{images/2764.png} 909 \\
 \hline
 4 & \includegraphics[height=1em]{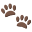} 1643 & \includegraphics[height=1em]{images/1f64f.png} 1193 & \includegraphics[height=1em]{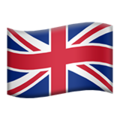} 2622 & \includegraphics[height=1em]{images/1f64f.png} 779 \\
 \hline
 5 & \includegraphics[height=1em]{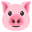} 1530 & \includegraphics[height=1em]{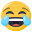} 823 & \includegraphics[height=1em]{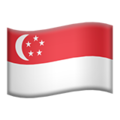} 2581 & \includegraphics[height=1em]{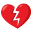} 760 \\
 \hline
 6  &  \includegraphics[height=1em]{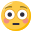} 1034 &  \includegraphics[height=1em]{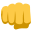} 794  & \includegraphics[height=1em]{images/1f64f.png} 2225 &  \includegraphics[height=1em]{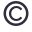} 616\\
 \hline
 7  & \includegraphics[height=1em]{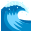} 934 & \includegraphics[height=1em]{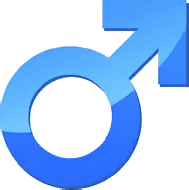} 726 & \includegraphics[height=1em]{images/2764.png} 1702 & \includegraphics[height=1em]{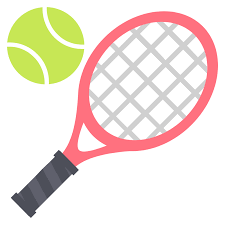} 513\\
 \hline
 8  & \includegraphics[height=1em]{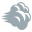} 820 & \includegraphics[height=1em]{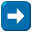} 725 & \includegraphics[height=1em]{images/1f494.png} 386 & \includegraphics[height=1em]{images/1f602.png} 510 \\
 \hline
 9  & \includegraphics[height=1em]{images/1f1fa-1f1f8_rect.png}  625 & \includegraphics[height=1em]{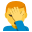} 724  &  \includegraphics[height=1em]{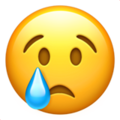} 340 & \includegraphics[height=1em]{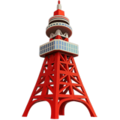} 433\\
 \hline
 10  & \includegraphics[height=1em]{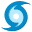} 367 & \includegraphics[height=1em]{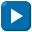} 683 & \includegraphics[height=1em]{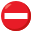} 259 & \includegraphics[height=1em]{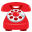} 361\\ [1ex] 
 \hline
\end{tabular}
\end{table}

To address RQ2, we show in Table \ref{emoji_ranks} the top ten most frequently used emojis across both crisis events in the tweets that express solidarity and those that do not. We observe that \includegraphics[height=1em]{images/2764.png} is used more frequently in the Irma solidarity tweets (Rank 3) but not in the Irma tweets that do not express solidarity. In the top 10 Irma emojis used in tweets not expressing solidarity, we also observe more negatively valenced emojis, including \includegraphics[height=1em]{images/1f62d.png} and \includegraphics[height=1em]{images/1F629.png}. The \includegraphics[height=1em]{images/1f602.png} emoji is interesting, since the prevailing meaning is ``face with tears of joy'', however this emoji can sometimes be used to express sadness \cite{wijeratne2016emojinet}. In addition,  \includegraphics[height=1em]{images/1f64f.png} is used across all four sets, albeit at different ranks (e.g. Rank 1 in Irma solidarity and Rank 6 in Paris solidarity tweets). 

When comparing the two crisis events, we make the observation that the top 5 ranked Paris solidarity emojis are flags of different countries, related to expressions of solidarity from these countries, including France (\includegraphics[height=1em]{images/1f1eb-1f1f7_rect.png}) at Rank 1, while  \includegraphics[height=1em]{images/1f1fa-1f1f8_rect.png} appears at Rank 9 in the Irma solidarity set. We can thus observe that even though the underlying behavior we study in these two events is solidarity, the top emojis used to express such behavior are different in the two events. During the Paris event, solidarity is signaled through the use of flag emojis from different countries, while in the Irma corpus flag emojis do not play a prominent role.  

\begin{figure}[t]
    \includegraphics [width=8cm, height=6.5cm]{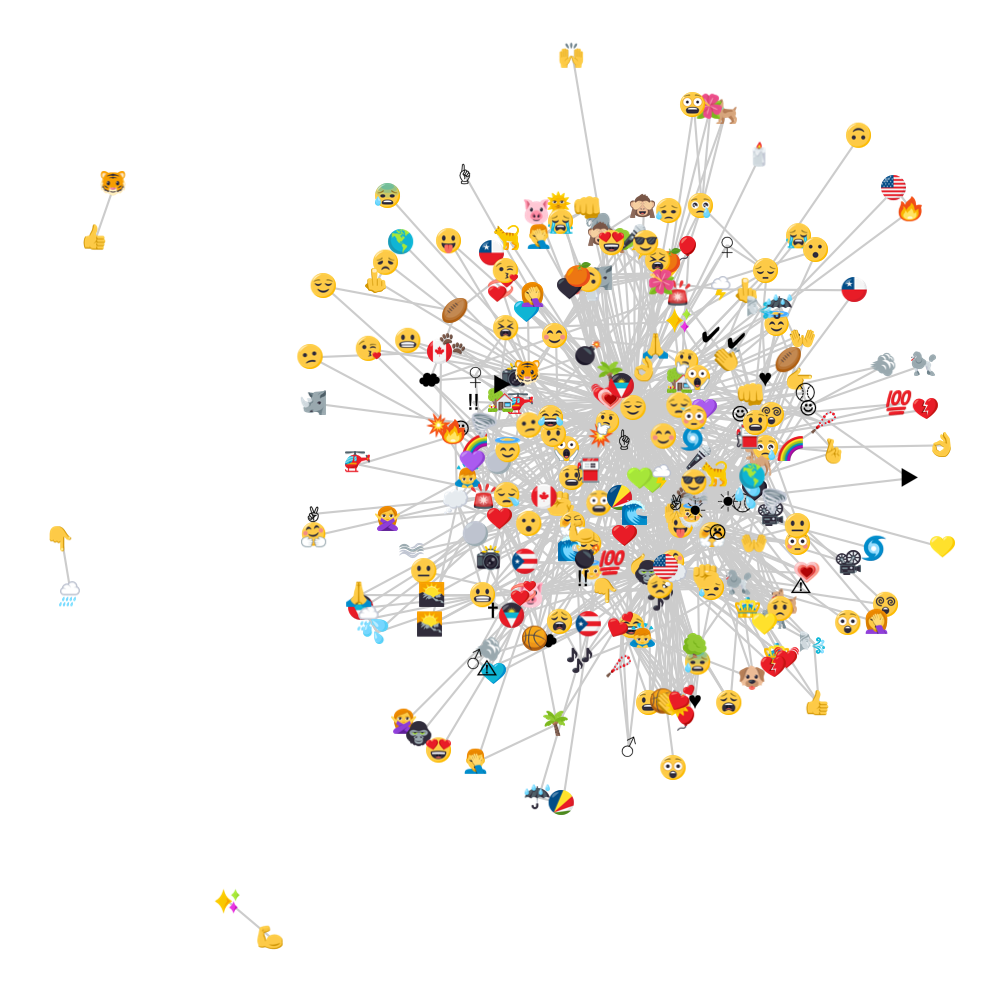}
  \caption{Cooccurrence network for emojis expressing solidarity from regions affected by Hurricane Irma}
  \label{fig:emoji_cooccurrence_irma_US}
\end{figure}

\textbf{RQ3: Which emojis coocur in tweets that are posted within areas directly affected by crisis events as compared to those tweets that are posted from other areas?}

This research question and the two following RQs are driven by the hypothesis that solidarity would be expressed differently by people that are directly affected by the crisis than those who are not \cite{buechler2016understanding}.
 To address RQ3, we first geotagged tweets using {\fontfamily{qcr}\selectfont{geopy}} Python geocoding library\footnote{https://github.com/geopy/geopy} to map the users' locations to their corresponding country. Table \ref{percentage} shows the total number of emojis in solidarity tweets that were geotagged and categorized as posted within regions affected by the event vs. other regions. We then built co-occurrence networks of emojis in both Irma and Paris corpora using the R {\fontfamily{qcr}\selectfont{ggnetwork}} package\footnote{https://tinyurl.com/y7xnw9lr} with the force-directed layout to compare these emoji co-occurrence networks in solidarity tweets that were posted within areas directly affected by the crisis and the areas that were not (shown in Figures 1-4).

\begin{figure}[t]
    \includegraphics[width=8cm, height=6.5cm]{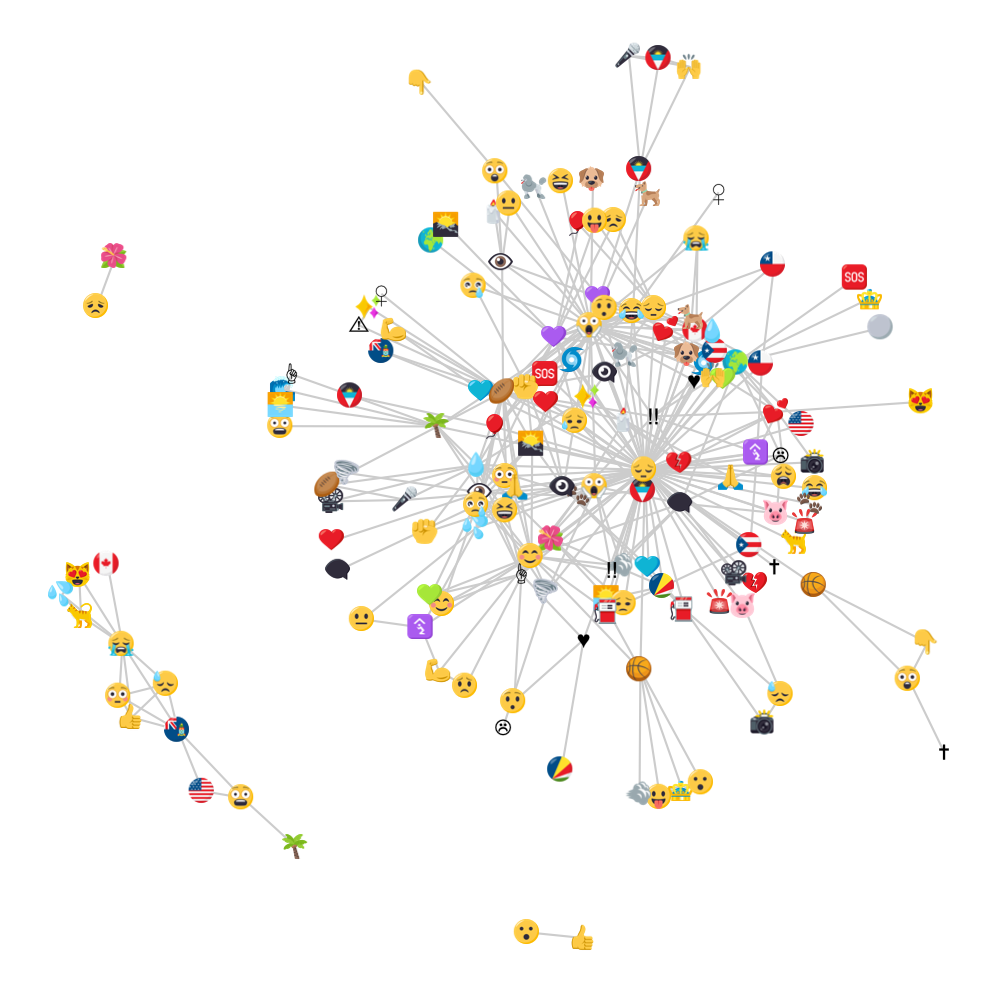}
  \caption{Cooccurrence network for emojis expressing solidarity from regions \textbf{not} affected by Hurricane Irma}
  \label{fig:emoji_coocurrence_irma_notUS}
\end{figure}

\begin{table}[h]
\centering
\caption{Total count and proportion of emojis in geotagged tweets from affected vs. other regions} \label{percentage}
\begin{tabular}{|l|l|l|l|l|}
\hline
 & \textbf{Irma}    & \textbf{Paris}  \\ \hline
Affected Regions       & 10048 (67.81\%)         & 925 (6.52\%) \\ \hline
Other Regions        & 4770 (32.19\%)&    13267 (93.48\%)       \\ \hline
\end{tabular}
\end{table}

Figure \ref{fig:emoji_cooccurrence_irma_US} represents the co-occurrence network of emojis within the regions affected by the Hurricane Irma (United States, Antigua and Barbuda, Saint Martin, Saint Barthelemy, Anguilla, Saint Kitts and Nevis. Birgin Islands, Dominican Republic, Puerto Rico, Haiti, Turks and Caicos and Cuba)\footnote{http://www.bbc.com/news/world-us-canada-41175312}.

We find the  pair \includegraphics[height=1em]{images/1f64f.png} \textbf{\textendash}  \includegraphics[height=1em]{images/2764.png} occurs most frequently in solidarity tweets collected within the Irma affected regions. The other top co-occurring pairs following the sequence include \includegraphics[height=1em]{images/1f499.png} \textbf{\textendash}  \includegraphics[height=1em]{images/1f437.png}, \includegraphics[height=1em]{images/1f64f.png} \textbf{\textendash}  \includegraphics[height=1em]{images/1f1fa-1f1f8_rect.png}, \includegraphics[height=1em]{images/1f30a.png} \textbf{\textendash}  \includegraphics[height=1em]{images/1f4a8.png} and \includegraphics[height=1em]{images/1f4a8.png} \textbf{\textendash}  \includegraphics[height=1em]{images/1f633.png}; these pairs might convey the concerns expressed in the tweets that originate within affected areas. The \includegraphics[height=1em]{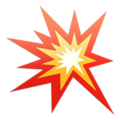} emoji appears at the centre of the network denoting the impact of the Irma event. The \includegraphics[height=1em]{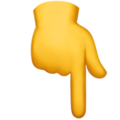} \textbf{\textendash}  \includegraphics[height=1em]{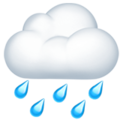}, \includegraphics[height=1em]{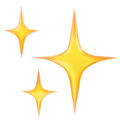} \textbf{\textendash}  \includegraphics[height=1em]{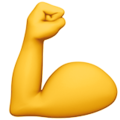}, \includegraphics[height=1em]{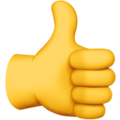} \textbf{\textendash}  \includegraphics[height=1em]{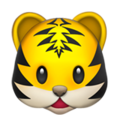} are the emojis that appear in isolation from the network. The \includegraphics[height=1em]{images/1f4aa.png} and \includegraphics[height=1em]{images/t1f42f.png} emojis can serve as indicators to stay strong during this hurricane calamity.

Figure 2 represents the co-occurrence network of emojis in tweets posted outside the regions affected by Hurricane Irma.
We find that the pair \includegraphics[height=1em]{images/1f499.png} \textbf{\textendash} \includegraphics[height=1em]{images/1f437.png} tops the co-occurrence list. Next, we have other co-occurring pairs like  \includegraphics[height=1em]{images/1f30a.png}
 \textbf{\textendash} \includegraphics[height=1em]{images/1f4a8.png} and \includegraphics[height=1em]{images/1f4a8.png} \textbf{\textendash} \includegraphics[height=1em]{images/1f633.png} following the top most frequent pair in sequence. 
  We see the three disjoint networks apart from the main co-occurrence network. The \includegraphics[height=1em]{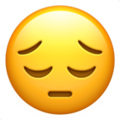} emoji appears  at the centre of the network expressing sorrow and the concern of the people during the event. The disjoint networks also contain flags and other emojis that express sadness and sorrow.

\begin{figure}[t]
    \includegraphics[width=8cm, height=6.5cm]{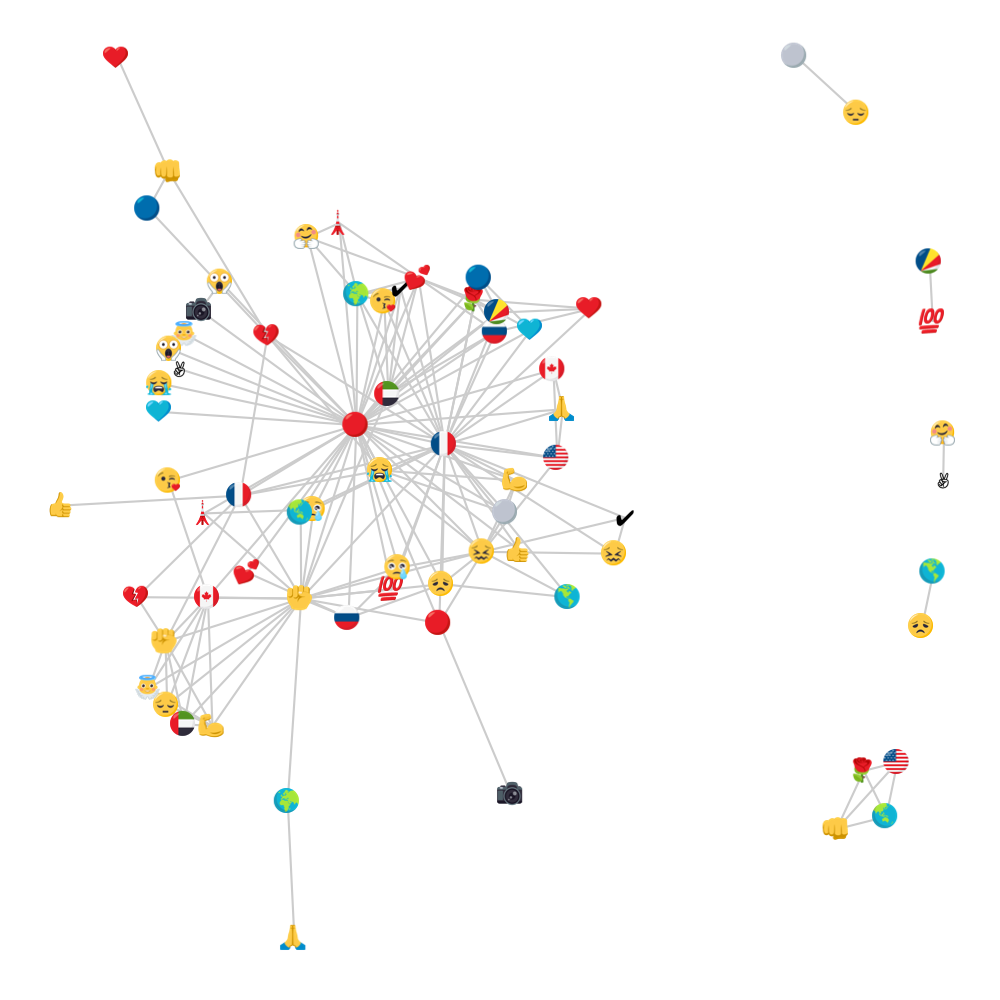}
  \caption{Cooccurrence network for emojis expressing solidarity from regions affected by November 2015 terrorist attacks in France}
  \label{fig:Emoji-Cooccurence_Paris}
\end{figure}

Figure 3 shows the co-occurrence network of emojis for November terrorist attacks in France. Within France, the pair \includegraphics[height=1em]{images/1f1eb-1f1f7_rect.png} \textbf{\textendash} \includegraphics[height=1em]{images/2764.png} tops all the co-occurrence pairs. Co-occurrence pairs like  \includegraphics[height=1em]{images/1f64f.png}
 \textbf{\textendash} \includegraphics[height=1em]{images/1f1eb-1f1f7_rect.png} and \includegraphics[height=1em]{images/1f64f.png}
 \textbf{\textendash} \includegraphics[height=1em]{images/2764.png} follow the top co-occurring sequence, strongly conveying the solidarity of people who tweeted during November terrorist attacks. We also have co-occurring pairs containing flags of other countries following the top-tweeted list that shows uniform feeling among the people by trying to express their sorrow and prayers. 
 The \includegraphics[height=1em]{images/1f534.png} emoji appears in the centre of the large network as an expression of danger during terrorist attacks. We can also see that the network contains many flags that indicates the concern and worries of people from many different countries. 
 There are five disjoint networks that again contain emojis that express the sorrow, prayers and discontent.
 
Figure 4 represents the co-occurrence network of emojis for November terrorist attacks in Paris outside France. We find that the pair \includegraphics[height=1em]{images/1f1eb-1f1f7_rect.png} \textbf{\textendash} \includegraphics[height=1em]{images/2764.png} tops the co-occurrence list as within France, which is followed by the co-occurring pairs \includegraphics[height=1em]{images/1f64f.png}
 \textbf{\textendash} \includegraphics[height=1em]{images/1f1eb-1f1f7_rect.png} and \includegraphics[height=1em]{images/1f64f.png}
\textbf{\textendash} \includegraphics[height=1em]{images/2764.png}. 
We can infer that the people within or outside France shared common emotions that includes a mixture of prayers, support and concern towards Paris and its people. 
We find the \includegraphics[height=1em]{images/1f1eb-1f1f7_rect.png}, \includegraphics[height=1em]{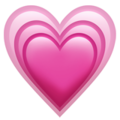} and \includegraphics[height=1em]{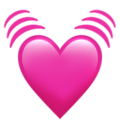} appear at the centre of the network to convey solidarity. The \includegraphics[height=1em]{images/1f4a5.png} emoji also appears at the center, similar to Figure \ref{fig:emoji_cooccurrence_irma_US}. One important inference is that the \includegraphics[height=1em]{images/1f4a5.png} emoji appears at the network center Irma affected regions whereas it appears at the network center for \textit{unaffected} regions in Paris event.

\begin{figure}[t]
    \includegraphics[width=8cm, height=6.5cm]{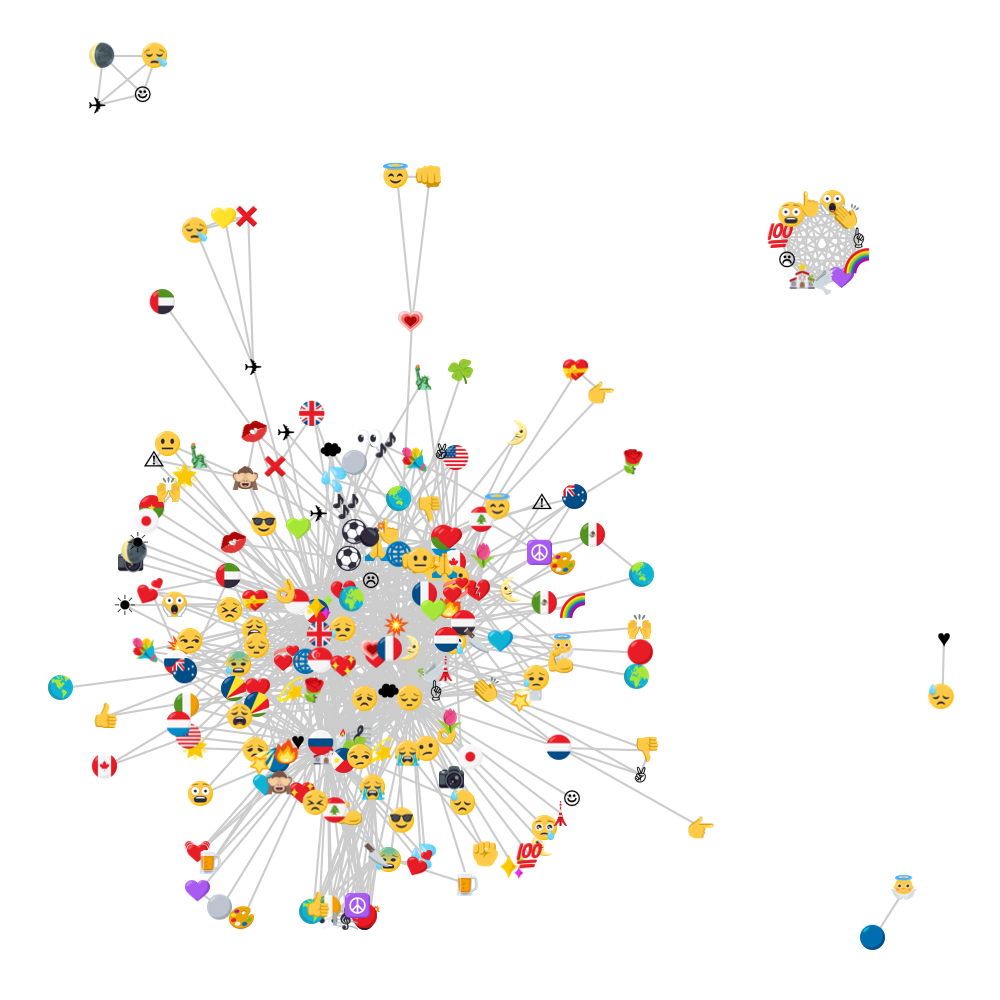}
  \caption{Cooccurrence network for emojis expressing solidarity from regions \textbf{not} affected by November 2015 terrorist attacks in France}
  \label{fig:Emoji-Cooccurence_NotParis}
\end{figure}

\textbf{RQ4: How can emojis be used to understand the diffusion of solidarity expressions over time?}

For addressing this research question, we plot in Figures 5 and 6 the diffusion of emojis across time (filtering emojis that occur fewer than 50 times and 25 times per day resp. for the 26197 emojis in Irma and 24801 emojis in the Paris solidarity corpus (c.f. Table \ref{emoji_descriptives}). The emojis are arranged on y-axis based on their sentiment score based on the publicly available work done by Novak et al. \cite{novak2015sentiment}.

\begin{figure}[h]
    \includegraphics[width=8cm, height=9cm]{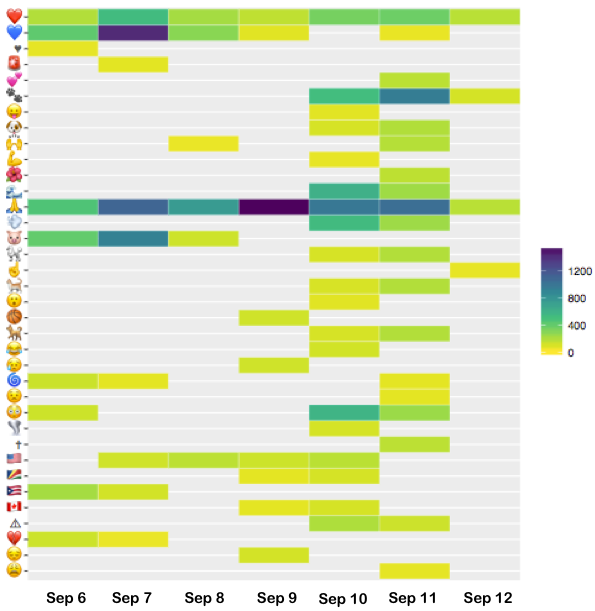}
  \caption{Diffusion of emojis across time for the Hurricane Irma disaster (N=26197 emojis)}
  \label{fig:diffusion_irma}
\end{figure}

\begin{figure}[h]
    \includegraphics[width=8cm, height=9cm]{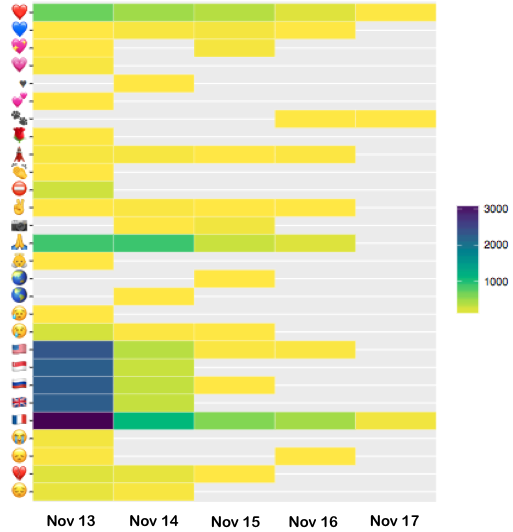}
  \caption{Diffusion of emojis across time for the November Paris attacks  (N=24801 emojis)}
  \label{fig:diffusion_paris}
\end{figure}

In the Irma corpus, the temporal diffusion of emojis is quite interesting (Figure \ref{fig:diffusion_irma}). Hurricane Irma grabbed attention of the world on September 6\textsuperscript{th} when it turned into a massive storm and the reaction on social media expressing solidarity for Puerto Rico was through \includegraphics[height=1em]{images/2764.png} and \includegraphics[height=1em]{images/1f64f.png}. During the following days, the United States is in the path of the storm, and there is an increased presence of \includegraphics[height=1em]{images/1f1fa-1f1f8_rect.png} and the presence of other countries flags. As the storm lashes out on the islands on September 7\textsuperscript{th}, people express their feelings through \includegraphics[height=1em]{images/2764.png} and \includegraphics[height=1em]{images/1f64f.png} emojis and also warn people about caring for the pets. As the storm moves through the Atlantic, more prayers with \includegraphics[height=1em]{images/2764.png} and \includegraphics[height=1em]{images/1f614.png} emojis emerge on social media for people affected and on the path of this storm. The storm strikes Cuba and part of Bahamas on September 9\textsuperscript{th} before heading towards the Florida coast. As the storm moves towards the US on September 10\textsuperscript{th}, people express their thoughts through \includegraphics[height=1em]{images/1f30a.png}, \includegraphics[height=1em]{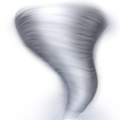} and \includegraphics[height=1em]{images/1f4a8.png} causing tornadoes. When images of massive flooding emerge on social media, people respond with pet emojis like \includegraphics[height=1em]{images/1f437.png}, \includegraphics[height=1em]{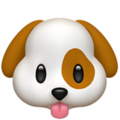}, \includegraphics[height=1em]{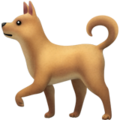} and \includegraphics[height=1em]{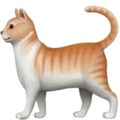} to save them. The \includegraphics[height=1em]{images/1f437.png} emoji may also serve as an indicator of high-pitched crying \cite{wijeratne2016emojinet}.

\begin{figure}[h]
    \includegraphics[width=8cm, height=8cm]{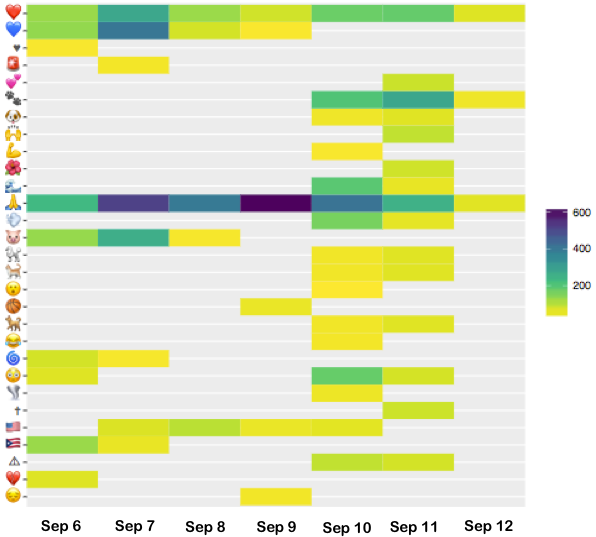}
  \caption{Diffusion of emojis across time for the Hurricane Irma within affected regions  (N=10048 emojis)}
  \label{fig:diffusion_irma_affected}
\end{figure}

\begin{figure}[h]
    \includegraphics[width=8cm, height=8cm]{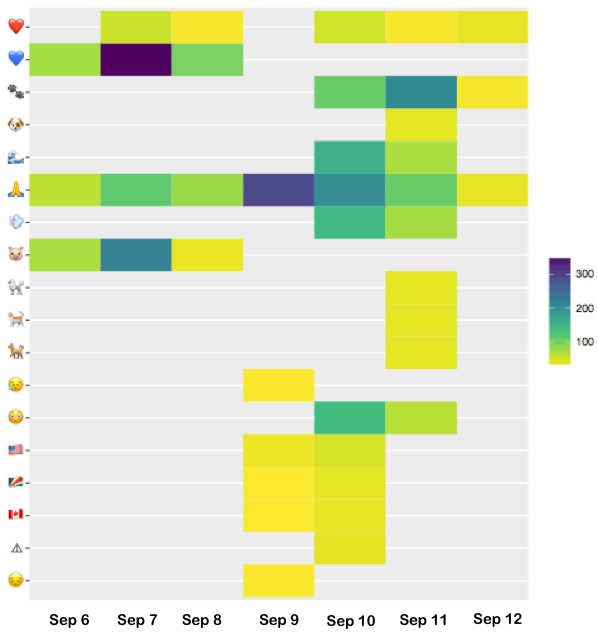}
  \caption{Diffusion of emojis across time for the Hurricane Irma outside affected regions  (N=4770 emojis)}
  \label{fig:diffusion_irma_outside}
\end{figure}

In the Paris attacks, the first 24 hours from 13\textsuperscript{th} night to 14\textsuperscript{th} were the days on which most number of emojis were used. When news of this horrible attack spreads on social media, the immediate reaction of the people was to express solidarity through hashtags attached with \includegraphics[height=1em]{images/1f1eb-1f1f7_rect.png} emoji. As a result,  \includegraphics[height=1em]{images/1f1eb-1f1f7_rect.png} was the most frequently used emoji across all days. Emojis of other country flags such as \includegraphics[height=1em]{images/1f1fa-1f1f8_rect.png}, \includegraphics[height=1em]{images/1f1ec-1f1e7_rect.png} emerge to indicate solidarity of people from these countries with France. Even after the end of the attack on Nov 13\textsuperscript{th}, people express prayers for the people of France through \includegraphics[height=1em]{images/1f64f.png} emoji. Images and videos of the attacks emerge on social media on Nov 14\textsuperscript{th} leading to the use of the \includegraphics[height=1em]{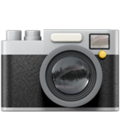} emoji. The \includegraphics[height=1em]{images/2764.png} emoji occurs across all days for the Paris event.

Across both events, we also observe a steady presence over all days of positively valenced emojis in the tweets expressing solidarity (the top parts of the diffusion graphs), while negatively valenced emojis are less prevalent over time (e.g. \includegraphics[height=1em]{images/1f494.png} appears in the first two and three days in the Irma and Paris events resp.).

\textbf{RQ5: How can emojis be used to study the temporal and geographical difussion of solidarity expressions during crisis events?}

Our primary aim with this research question was to look at how the emojis of solidarity diffuse over time within the affected community and compare communities not affected by the same event.
Using the geotagged tweets described in RQ2, we created Figures \ref{fig:diffusion_irma_affected} and \ref{fig:diffusion_irma_outside} to represent the diffusion of emojis over time within the affected regions on the path of Hurricane Irma and the non-affected regions respectively. Since the distribution of emojis in geotagged tweets for the Paris attacks is skewed (6.52\% emojis expressing solidarity from affected regions, c.f. Table \ref{percentage}), we have excluded the Paris event for analysis in this RQ.

Figures \ref{fig:diffusion_irma_affected} and \ref{fig:diffusion_irma_outside} allow us to contrast how the Hurricane Irma event is viewed within and outside the affected regions. On September 6\textsuperscript{th}, \includegraphics[height=1em]{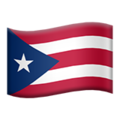} emerged when the hurricane started battering the islands along with a lot of heart emojis. On the other hand, more heart emojis emerged expressing solidarity from outside the affected regions when the people realized the effect of the hurricane (Sep 7\textsuperscript{th}. As the storm moved forward, people in the affected regions express a lot of prayers. On September 9\textsuperscript{th}, as the storm moves towards the United States after striking Cuba, there seems to be more prayers \includegraphics[height=1em]{images/1f64f.png} amongst affected as well as outside communities. The  \includegraphics[height=1em]{images/1f64f.png} emoji is constant in the affected regions. In both Figures \ref{fig:diffusion_irma_outside} and \ref{fig:diffusion_irma_affected}, we see that variety of emojis appear in the latter days of the event (starting Sep 9\textsuperscript{th}), including the \includegraphics[height=1em]{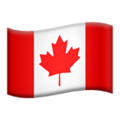}, \includegraphics[height=1em]{images/1f622.png}, and \includegraphics[height=1em]{images/1f633.png} as well as the animal/pet emojis, likely indicating the emergence of different topics of discourse related to the event.

\section{Conclusion and Future Work}
We described our data, algorithm and method to analyze corpora related to two major crisis events, specifically investigating how emojis are used to express solidarity on social media. Using manual annotation based on hashtags, which is a typical approach taken to distance label social media text from Twitter \cite{mintz2009distant}, we categorized tweets into those that express solidarity and tweets that do not express solidarity. We then analyzed how these tweets and the emojis within them diffused in social media over time and geographical locations to gain insights into how people reacted globally as the crisis events unfolded. We make the following overall observations:
\begin{itemize}
    \item Emojis are a reliable feature to use in classification algorithms to recognize expressions of solidarity (\textbf{RQ1}).
    \item The top emojis for the two crisis event reveal the differences in how people perceive these events; in the Paris attack tweets we find a notable presence of flag emojis, likely signaling nationalism but not in the Irma event (\textbf{RQ2}).
    \item Through the cooccurence networks, we observe that the emoji pairs in tweets that express solidarity include anthropomorphic emojis (\includegraphics[height=1em]{images/1f64f.png}, \includegraphics[height=1em]{images/1f614.png}, \includegraphics[height=1em]{images/2764.png}) with other categories of emojis such as \includegraphics[height=1em]{images/1f4a8.png} and \includegraphics[height=1em]{images/1f1eb-1f1f7_rect.png} (\textbf{RQ3}). 
    \item Through analyzing the temporal and geospatial diffusion of emojis in solidarity tweets, we observe a steady presence over all days of positively valenced emojis, while negatively valenced emojis become less prevalent over time (\textbf{RQ4, RQ5}).
\end{itemize}

\textbf{Future Work}: While this paper addressed five salient research questions related to solidarity and emojis, there are a few limitations. First, our dataset contains emojis that number in the few thousand, which is relatively small when compared to extant research in emoji usage \cite{ljubevsic2016global}. However, we aim to reproduce our findings on larger scale corpora in the future. Second, we analyzed solidarity during crisis events including a terrorist attack and a hurricane, whereas solidarity can be triggered without an overt shocking event, for example the \textit{\#MeToo} movement. 
In future work, there is great potential for further investigation of emoji diffusion across cultures. In addition to categorizing tweets as being posted from affected regions and outside of affected regions, we wish to analyze the geographical diffusion with more granularity, using countries and regions as our units of analysis to better understand the cultural diffusion of solidarity emojis. An additional future goal is to analyze the interaction of sentiment of emojis and solidarity as well as the text that cooccurs with these emojis in further detail. We anticipate our approach and findings will help foster research in the dynamics of online mobilization, especially in the event-specific and behavior-specific usage of emojis on social media.

\subsubsection*{Acknowledgements}
We are grateful for the extremely helpful and constructive feedback  given by anonymous reviewers. We thank the two trained annotators for their help in creating our dataset. This research was supported, in part, by a fellowship from the National Consortium of Data Science to the last author.

\bibliographystyle{alpha} 
\bibliography{emoji2018}
\end{document}